\documentstyle[epsf, rotate]{l-aa}

\def\lesssim{\lower4pt \hbox{$\buildrel < \over \sim$}}
\def\gtrsim{\lower4pt \hbox{$\buildrel > \over \sim$}}

\begin{document}

\title{Time-dependent photoelectric absorption, photoionization and 
fluorescence line emission in Gamma-ray burst environments}

\author{Markus B\"ottcher$^1$, Charles D. Dermer$^2$, 
Anthony W. Crider$^1$, and Edison P. Liang$^1$}

\institute{$^1$Department of Space Physics and Astronomy, Rice University, 
6100 S. Main St., Houston, TX 77005-1892 \\
$^2$E. O. Hulburt Center for Space Research, Code 7653,
Naval Research Laboratory, Washington, DC 20375-5352}

\date{Received; accepted}
\offprints{M. B\"ottcher}
\thesaurus{02.01.4; 02.18.7; 13.07.1; 13.25.1}

\maketitle
\markboth{M. B\"ottcher et al.: Photoelectric absorption in GRB environments}
{M. B\"ottcher et al.: Photoelectric absorption in GRB environments}

\begin{abstract} 
If $\gamma$-ray bursts are associated with dense star-forming 
regions in young galaxies, photoelectric absorption by the dense 
circumburster material (CBM) will occur. As the burst evolves, the 
surrounding material is photoionized, leading to fluorescence
line emission and reduced photoelectric absorption opacity. We have
analyzed this process in detail, accounting  for the time-dependent
photoelectric absorption, photoionization  and fluorescence line emission
from the CBM. We find that even if  GRBs are hosted in dense star-forming
regions, photoionization of  the GRB environment leads to a constant, but
very weak level of delayed fluorescence line emission on
timescales of weeks to years  after the burst. A temporally evolving
iron~K edge absorption feature can serve as diagnostic tool to reveal the
density structure of the  CBM and may provide an opportunity for redshift
measurements. We also investigated whether photoelectric absorption could be 
responsible for the spectral evolution of the low-energy slopes 
of some bright BATSE $\gamma$-ray bursts displaying extremely 
hard spectra below the peak energy, inconsistent with the 
optically-thin synchrotron shock model. We find that a very 
strong metal enrichment ($\sim 100$ times solar-system abundances) 
in the $\gamma$-ray burst environment and a rather peculiar spatial
distribution of the CBM would be necessary in order to account 
for the observed hard spectra below a few 100~keV and their 
temporal evolution.
\keywords{Atomic processes --- Radiative transfer --- Gamma rays: bursts
--- X-rays: bursts}
\end{abstract}

\section{Introduction}

The recent discovery of X-ray and optical afterglows of
$\gamma$-ray bursts as a result of the Italian-Dutch 
BepoSAX mission (e. g., Costa et al. 1997;  Metzger et 
al. 1997; van Paradijs et al. 1997; Kulkarni et al. 1998)
has given strong support for the cosmological blastwave
model for $\gamma$-ray bursts (Rees \& M\'esz\'aros 
1992; M\'esz\'aros \& Rees 1993; Katz 1994). The observed 
temporal decay of the afterglow emission, generally well 
described by a power-law $F_{\nu} (t) \propto t^{-\alpha}$ 
with $1.1 \lesssim \alpha \lesssim 1.5$ (Costa et al.
1997; Feroci et al. 1998; Piro et al. 1998) and the overall
spectral shape of the afterglow emission are in very
good agreement with optically thin synchrotron emission 
by relativistic electrons accelerated at a relativistic 
blast wave as it expands and sweeps up matter from the 
surrounding medium (e.\ g., Katz 1994, Tavani 1996, 
Wijers, Rees \& M\'esz\'aros 1997, Dermer \& Chiang 1998).

If $\gamma$-ray bursts are associated with the collapse of
massive objects (e.\ g., Paczy\'nski 1998), they
are expected to be correlated with dense star-forming regions
and are possibly located within them. Recent comparisons of the
GRB flux distribution with the star-formation history of the
early universe seem to support this hypothesis (Wijers et al. 
1998). If this is true, photoelectric absorption,
photoionization and fluorescence line emission in the vicinity
of $\gamma$-ray bursts will inevitably occur and might lead to
observable consequences. Recently, M\'esz\'aros \& Rees 
(1998) have qualitatively discussed absorption 
edges and line features from atoms irradiated by a blastwave 
evolving in the nonradiative regime, for both a low denstiy interstellar 
medium or a dense burst environment which might appropriate for the
hypernova  scenario (Paczy\'nski 1998). They suggest that for 
reasonable parameter values absorption edges as well as fluorescence
lines and resonance-scattered burst emission could be detectable 
with future, rapid follow-up observations of GRB afterglows. 
They also propose that the non-detection of optical afterglows
for some bursts for which X-ray afterglows have been observed
and precisely located by the BeppoSAX satellite, could be due
to photoelectric absorption and extinction by gas and dust in the GRB
environment. Ghisellini  et al. (1998) have discussed the detectability  
of the iron K edge and K$\alpha$ fluorescence line in this 
context and suggested that it might provide a possibility 
to measure the redshift of GRBs. They also predict that the 
Fe~K$\alpha$ line should still be detectable long after the 
continuum X-ray afterglow has faded away. 

The process of time-dependent photoionization and photoelectric 
absorption has been studied in detail by  Perna \& Loeb 
(1998). They focus on optical absorption
lines originating in the CBM and point out the effect that 
photoionization of the GRB environment by the burst radiation 
leads to a reduction of the absorption
opacity,  and suggest the temporal evolution of absorption line 
shapes and equivalent widths as a new diagnostic tool to 
map the GRB environment. This method is applicable in 
the case of very dense GRB environments, in which case the
evolution of the absorption line shapes occurs on a sufficiently
short timescale to be measurable within the short periods during
which optical GRB afterglows remain observable. 

In this paper, we focus on the observable effects in the X-ray
regime of the time-dependent photoelectric absorption, photoionization 
and fluorescence line emission processes in the vicinity of a 
cosmological $\gamma$-ray burst. We adopt a similar approach
to the one used by Perna \& Loeb (1998), but
additionally implementing the Auger process and
fluorescence line emission following inner-shell ionization
events. We calculate the temporal evolution of the absorbed 
burst spectrum, present light curves of the continuum and 
fluorescence line emission, and suggest observational tests 
to reveal the density structure of GRB environments on the 
basis of GRB follow-up X-ray observations.

Another motivation for this work is the finding (Crider et
al. 1997; Preece  et al. 1998) that
the temporally resolved low-energy spectra of many $\gamma$-ray bursts
during the early phases of the burst are inconsistent with optically thin
synchrotron emission from ultrarelativistic electrons. This latter
process gives  an energy spectral index ($F_{\nu} \propto
\nu^{\alpha}$) 
$\alpha = 1/3$ if electrons in the blast wave are inefficiently 
cooled, maintaining an ultrarelativistic low-energy cutoff 
(Katz 1994) and $\alpha = -1/2$ if synchrotron 
cooling is efficient, yielding an electron power-law spectrum 
in energy with index $p = - 2$. A significant fraction of 
snap-shot burst spectra have differential slopes harder than 1/3 at 
the low-energy end of sensitivity range of the BATSE instrument 
(Preece et al. 1998), and in these bursts the 
low-energy spectra generally evolve from hard to soft during the 
decay phase of the burst radiation (Crider et al. 1997). 
We investigated whether photoelectric absorption of the intrinsic 
$\gamma$-ray burst spectrum in the vicinity of the GRB could
be responsible for the hard low-energy spectra of GRBs and their
temporal evolution. This process has first been studied in
the context of the low-energy spectra of $\gamma$-ray bursts by
Liang \& Kargatis (1994). Considering the photoelectric
absorption of an unbroken power-law spectrum by pure neutral iron 
along the line of sight to a GRB, they found that an absorption 
depth at the iron K edge of several hundred is necessary to account 
for the steep $\gamma$-ray burst spectra if the intrinsic spectrum 
is an unbroken power-law with the spectral index observed above the 
break. Assuming a solar-system abundance of iron in the absorbing 
matter, this corresponds to a Thomson depth of $\tau_T \gtrsim 100$, 
which would result in strong distortion of the high-energy spectrum 
due to Compton downscattering. They therefore concluded that this 
scenario may be ruled out. Using our detailed photoionization
calculations, we have re-investigated this idea under more 
realistic assumptions, and basically confirm the previous
results of Liang \& Kargatis (1994).

In Section~2, we describe the model assumptions and the computational
procedure which we use to solve the time-dependent photoionization
and radiative transfer problem. We discuss the expected evolution of
photoelectric absorption features and the luminosity and light 
curves of fluorescence lines under various assumptions for
the density in the GRB environment in Section~3. In Section~4 we 
present a short discussion on the application of our code to the
low-energy spectra of time-resolved BATSE GRB spectra. We summarize 
and present our conclusions in Section~5.

\section{Model assumptions and computational scheme}

We simulate a relativistic blast wave expanding into and irradiating
and photoionizing a stationary external medium. The evolution of the 
blast wave and of the spectrum radiated by it is represented by the 
analytic parametrization of Dermer et al. (1998). The spectral 
evolution of our intrinsic model GRB spectra is illustrated 
in Fig. 1. 

\begin{figure}
\epsfysize=6cm
\rotate[r]{
\epsffile[50 70 550 500]{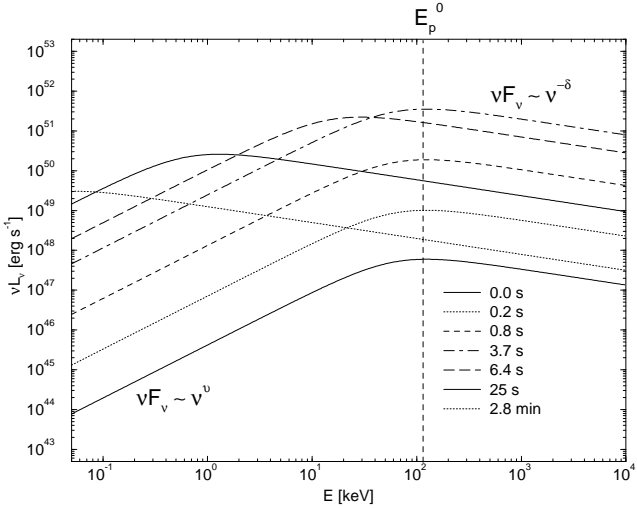}}
\caption[]{Evolution of the intrinsic (unabsorbed) model GRB 
spectrum. Parameters: $E_0 = 10^{54}$~erg, $\Gamma_0 = 300$, 
$g = 1.6$, $\eta = 0$, $\upsilon = 4/3$, $\delta = 0.4$.}
\end{figure}

The model is determined by the following parameters: 
$E_0 = 10^{54} \, E_{54}$~erg is the total energy transferred to 
relativistic baryons during the initial explosion, and $\Gamma_0$ is 
the initial bulk Lorentz factor of the ejecta forming the 
relativistic blast wave. The term $\upsilon$ is the spectral index
of the $\nu F_{\nu}$ spectrum below the spectral break (we use
$\upsilon = 4/3$, corresponding to optically thin synchrotron 
emission from a relativistic electron population), and $\delta$ is 
the spectral index above the break (we use $\delta = 0.4$ in 
our calculations; our final results are insensitive to this value).
The radius of the blast wave at a given time is denoted by
$r_b$. As the blast wave moves into the external medium, it 
sweeps up external matter and starts to decelerate when the 
energy of swept-up external matter in the rest frame of the
blast wave equals the rest-mass energy of the initial ejecta. 
The radius at which this happens is denoted $r_d$, the deceleration 
radius. Beyond this point, the bulk Lorentz factor $\Gamma$ of 
the blast wave decreases according to a power-law in radius,
determined by the deceleration index $g$, i.\ e.\ $\Gamma$ 
evolves as

\begin{equation}
\Gamma (r_b) = \cases{\Gamma_0 & for $r_b < r_d$ \cr
\Gamma_0 \, \left( {r_b \over r_d} \right)^{-g} & for $r_d \le r_b \le
r_d \, \Gamma_0^{1/g}$ \cr}.
\end{equation}
The CBM is assumed to have a power-law
profile in distance from the burst, determined by the
density $n_0$ at the deceleration radius and the power-law
index $\eta$:

\begin{equation}
n_{ext} (r) = n_0 \, \left( {r \over r_d} \right)^{-\eta}.
\end{equation}

The abundances of atoms and ions of the various elements are 
characterized by the abundance coefficients $X_a^i (r, t)$, 
where the subscript $a$ characterizes the element, and the
superscript $i$ characterizes the ionization state. The density of a
given element $a$ in ionization state $i$ is given by $n_a^i (r, t) 
= n_{ext} (r) \, X_a^i (r, t)$. Initially, we assume all elements 
in the external medium to be neutral and no metallicity gradients, though
the element abundances are allowed to differ from standard solar-system
abundances. In our calculations, we include H, He, C, N, O, Ne, Mg, Si,
S, Ar, Ca, Fe, and Ni, and neglect other elements. 

We start our simulation at $r_b \ll r_d$, i.\ e.\ at a point in time 
where the flux of the gamma-ray burst radiation is yet far below 
its maximum value. We split the external medium up into a radial
grid with steps $\Delta r$. Within each radial zone we 
calculate the absorption optical depth due to photoionization,

\begin{equation}
\Delta \tau_{abs} (r, E) = \Delta r \, n_{ext} (r) \sum\limits_{a, i}
X_a^i (r) \, \sigma_a^i (E),
\end{equation}
where the photoionization cross sections $\sigma_a^i (E)$ of all atoms 
and ions are evaluated using the relevant subroutines of the XSTAR
code (Kallman \& Krolik 1998; Kallman \& McCray 1982). The atoms and 
ions in each radial zone will be photoionized by the incident radiation. 
As the solution is advanced by a time step $\Delta t$ in the stationary 
frame, it will take a time interval $\Delta t_{rec} = \Delta t \, (1 - B)$, 
where $B = \sqrt{1 - 1/\Gamma^2}$, until the radiation from the blast wave 
emitted at the next time step reaches a fixed point. Thus, we assume that 
the zone located at radius $r$ is illuminated for a time step $\Delta 
t_{rec}$ by the constant incident spectrum $F_E (r, E) = L_E (r, E) \, 
e^{-\tau_{abs} (r, E)} / (4 \pi \, r^2)$. Each ionization state 
will be depopulated as

\begin{equation}
\left( {d X_a^i (r, t) \over dt} \right)_{-} = - X_a^i (r, t) \> 
\int dE \> {F_E (r, E) \over E} \sigma_a^i (E).
\end{equation}
Radiative transitions following a photoionization event will result
in a large variety of fluorescence lines. We include 200 strong
fluorescence lines from N, O, Ne, Mg, Si, S, Ar, Ca, Fe, and Ni,
using the ionization-state dependent fluorescence yields and 
line energies given in Kaastra \& Mewe (1993).  
We assume that all fluorescence lines are emitted isotropically at 
the location of the photoionization event. This leads to temporally 
delayed fluorescence line emission as seen by the observer: The 
fluorescence line radiation from an atom or ion located at a 
distance $r$ from the center, at an angle $\theta$ relative 
to the line of sight to the observer, will be observed 
$\Delta t_{fl} = (1 - \cos\theta) \, r/c$ later than the 
spherical light front which has photoionized the atom or 
ion.

Due to the high probability of Auger transitions following 
a K- or L-shell ionization of heavy atoms or ions, each 
photoionization event may result in effectively ejecting 
several electrons from the atom or ion. For this reason, e.\ g., 
it takes on average typically only 12, rather than 26 hard 
X-ray photons to ionize a neutral Fe atom completely. This 
is accounted for by means of a transition probability matrix, 
describing the probability $P_a^{i, j}$ of transition from 
ionization state $i$ to any higher ionization state $j$ after 
photoionization of an ion $(a, i)$. (Without the Auger process, 
$P_a^{i, j} = \delta_{i+1, j}$.) The evaluation of this
probability matrix requires the knowledge of the probability of
a K-, L(s)-, and L(p)-shell photoionization, which is calculated 
using the respective (sub)shell photoionization cross sections,
and the probabilities for an Auger transition following the
respective ionization events, as a function of the number of
ejected Auger electrons. For the elements with nuclear charge
$Z \le 18$, approximate values of these probabilities can be 
found in Weisheit (1974), while for the 
higher-Z elements the fluorescence and Auger yields are taken
from Kaastra \& Mewe (1993).

The higher ionization states ($j > 1$) will be populated due to 
ionization of lower-ionized ions according to

\begin{equation}
\left( {d X_a^j (r, t) \over dt} \right)_{+} = \sum\limits_{i < j} 
P_a^{i, j} \, X_a^i \, \int dE \> {F_E (r, E) \over E} \sigma_a^i (E).
\end{equation}
Since the radiation field of the GRB is very intense and the
density of surrounding material is assumed to be low ($n \lesssim
10^6 \, {\rm cm}^{-3}$), recombination may be neglected in our 
simulations. We also neglect the effect of resonant scattering
of fluorescence line emission because we are mostly interested
in inner-shell fluorescence lines (in particular the K$\alpha$
lines) which are generally non-resonant.

We solve iteratively the system consisting of eqs.\ (4), (5), 
and the exponential depletion of the incident radiation field 
due to photoelectric absorption. This scheme is propagated 
through all radial zones until the outer boundery of our model 
system is reached and the radiation is assumed to escape freely 
towards the observer. After each time step, the solution is 
forwarded a time step $\Delta t$, in which the blast wave moves 
out a distance $\Delta r_b = c \, B \, \Delta t$. In the receiving 
(observer's) frame, this corresponds to a time step $\Delta t_{rec} 
= \Delta t \, (1 - B)$. The new $r_b$ is used to re-calculate the 
burst emission, and the emitted spectrum is processed through the 
volume still located outside $r_b$ according to the scheme described 
above. 

\section{Absorption edges and delayed fluorescence line emission}

A fraction $Y$, the fluorescence yield, of all K-shell ionization
events of a certain element will be followed by a radiative transition,
resulting in the emission of a fluorescence line photon. For the 
heavy elements Ca, Fe and Ni, the K$\alpha$ transition is the most
probable radiative transition following a K-shell ionization.
As mentioned in the previous section, fluorescence line emission
will be observable from virtually all parts of the GRB
environment, while only a small solid angle $\sim 1/\Gamma^2$
of the blast wave contributes significantly to the observed
radiation as long as the blast wave is highly relativistic. 
This leads to a time delay between the direct burst radiation
and the fluorescence line emission from misaligned directions
of the burst environment, which will contribute most of the
fluorescence line fluence resulting from the burst if, as
we are assuming throughout this paper, the GRB emission is
uncollimated. 

The maximum time delay of this fluorescence line emission may be 
estimated from the size of the ionization sphere of the 
GRB, which is the region where a certain element is essentially
completely ionized. In a dense environment (which we call the optically
thick limit), the ionization radius is determined by the supply
of photons with energies above the ionization threshold of
the element under consideration. This radius is found by multiplying
the number 
\begin{equation}
N_a = X_a \; \int\limits_0^{r_{ion}} d\tilde r \> 4\pi \, {\tilde
r}^2 n_{ext}  (\tilde r) \, 
\end{equation}
of atoms or ions in the ionization sphere
 by  the number of photons $N_{a,ion}$ required to completely ionize an
atom, and equating this result with the number of ionizing photons 
emitted during the burst.  We obtain
\begin{equation}
r_{ion} ({\rm opt. \> thick}) = \left( {3 \, k \, E_0 \, \epsilon_K^{1 - w}
\over 4 \pi \, N_{a, ion} \, m_e \, X_a \, n_0 \, (w - 1)} \right)^{1/3},
\end{equation}
where $k$ is a normalization factor, 

\begin{equation}
k = \cases{0.33 \, \epsilon_p^{w - 2} & in the non-radiative limit ($g = 3/2$), 
\cr
0.30 \, \epsilon_p^{w - 2} & in the radiative limit ($g = 3$), \cr}
\end{equation}
$\epsilon_p = E_p^0 / (m_e c^2)$ is the normalized photon energy 
of the peak of the $\nu F_{\nu}$ spectrum at the peak of the burst 
light curve, $\epsilon_K = E_K / (m_e c^2)$ is the normalized K-edge 
energy, and $w = 3 (2 g + 1) / (4 g + \eta/2)$. Here, we have assumed
that the K-edge energy is much smaller than the peak of the $\nu F_{\nu}$
spectrum of the burst in the early phase of the burst, i.\ e.\ $\epsilon_K
\ll \epsilon_p$.

In a dilute environment (which we call the optically thin limit), 
the ionization radius is determined by the probability for an atom 
in the CBM to be ionized by a photon of the GRB radiation, which 
translates to the condition

\begin{equation}
\int\limits_{\epsilon_K}^{\infty} d\epsilon \> \sigma_a  (\epsilon)
\, {N_{ph} (\epsilon) \over 4 \pi r_{ion}^2} = 1.
\end{equation}
This yields

\begin{equation}
r_{ion} ({\rm opt. \> thin}) = \left( {k \, E_0 \, A_0 \, \epsilon_K^{1 - w}
\over 4 \pi \, m_e \, Z^2 \, [2 + w]} \right)^{1/2},
\end{equation}
where $Z$ is the nuclear charge of an atom/ion of species $a$ and 
we have approximated the photoionization cross section by $\sigma_a 
(\epsilon) \approx (A_0 / Z^2) \, (\epsilon / \epsilon_K)^{-3}$, 
where $A_0 = 6.3 \cdot 10^{-18}$~cm$^2$ and $\epsilon_K \approx 
2.66 \cdot 10^{-5} \, Z^2$.

\begin{figure}
\epsfysize=7cm
\epsffile[20 90 420 550]{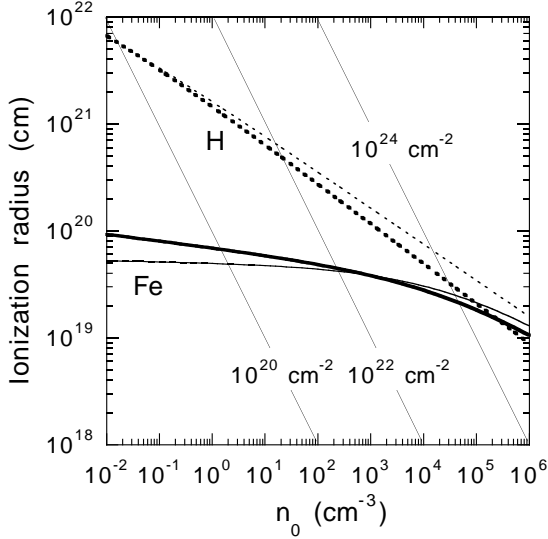}
\caption[]{The delay timescale of fluorescence line emission of iron
and hydrogen ionized by a GRB evolving in the radiative limit (thick
curves) and in the non-radiative limit (light curves), respectively,
in a uniform external medium as a function of external density $n_0$. 
Parameters: $\Gamma_0 = 300$, $\upsilon = 4/3$, $\delta = 0.4$, 
$E_0 = 10^{54}$~erg, standard solar-system iron abundance.}
\end{figure}

The delay timescale for fluorescence line emission is given
by $\tau_{fl} = r_{ion}/c$. The expressions for the optically 
thin and optically thick limits are joined smoothly. In Fig. 2, 
the size of the ionization  radius of hydrogen and iron is 
plotted as a function of  the external density for standard 
parameters with the solar-system iron abundance of $X_{Fe} =
3.16 \cdot 10^{-5}$. The figure demonstrates that extremely 
long afterglow durations are expected from fluorescence line 
emission. The actual duration of the afterglow will generally 
be determined by the physical size over which the CBM has a 
substantial density, rather than by the supply of ionizing 
photons from the $\gamma$-ray burst. This corresponds to an 
expected fluorescence line afterglow duration of order months 
to years even if the GRB is associated with a dense star-forming
region.

\begin{figure}
\epsfysize=6cm
\rotate[r]{
\epsffile[50 70 550 500]{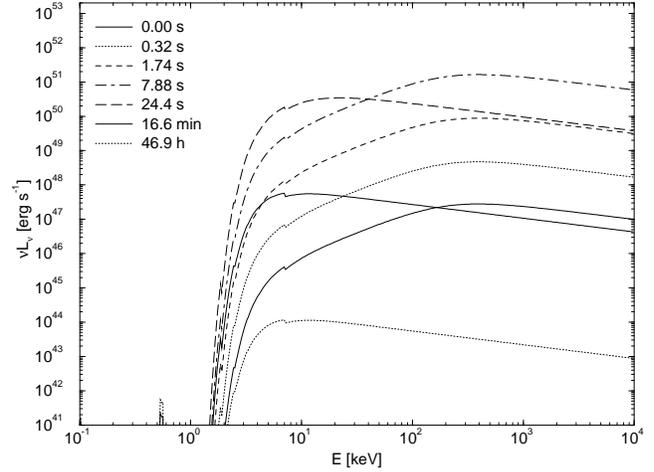}}
\caption[]{Temporal evolution of the observed GRB spectra at various 
times after the initial explosion. Parameters: $n_0 = 10$~cm$^{-3}$,
$r_{max} = 10$~kpc, $\Gamma_0 = 300$, standard solar-system element 
abundances.}
\end{figure}

The flux in fluorescence lines during this afterglow phase
is expected to remain roughly constant over the entire
duration of the fluorescence line afterglow because the
volume from which fluorescence line emission can be observed
at any given time scales as $\Delta V \propto r^2 \, \Delta r
\propto t^2 \, \Delta t$, where $\Delta t$ is the duration of 
the prompt GRB emission. This volume is illuminated by a 
continuum flux which scales as $F \propto r^{-2} \propto 
t^{-2}$ (only the outermost light fronts will be significantly 
affected by photoelectric absorption; thus, the illuminating 
fluence spectrum is basically the intrinsic burst spectrum), 
and a constant fraction $A$ of this flux is absorbed and re-radiated 
in fluorescence lines. Thus, $L_{fl} \propto A \, F \, \Delta V
\approx const.$

\begin{figure}
\epsfysize=6cm
\rotate[r]{
\epsffile[50 70 550 500]{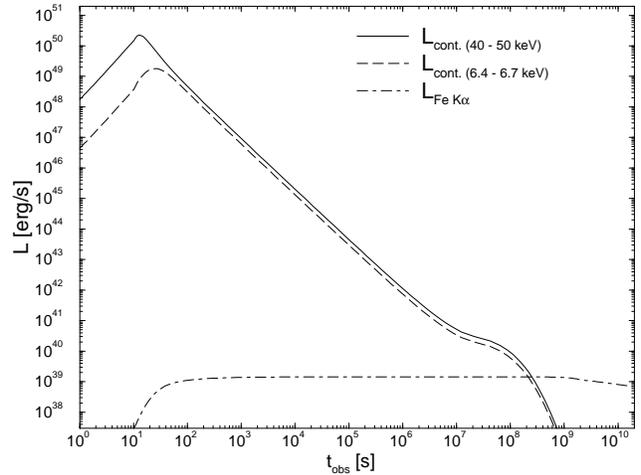}}
\caption[]{Energy flux light curves in the iron K$\alpha$ line band 
and in the hard X-ray continuum for the GRB spectral evolution shown 
in Fig. 2.}
\end{figure}

Figs. 3 and 4 demonstrate the results of numerical simulations for 
the case of a dilute, uniform medium with $n_0 = 10$~cm$^{-3}$
extending out to a distance of $r_{max} = 10$~kpc from the location
of the explosion. This might be representative of the case that the
GRB is hosted within a galaxy, but not directly inside a star-forming
region. Here we use the standard parameters for the intrinsic
GRB spectrum as quoted in the previous section. Fig. 3 shows the 
observable GRB spectra for various observing times. An iron 
absorption edge around 7~keV is visible. Even in the late afterglow 
phase the burst spectrum remains heavily absorbed below a few keV. 
On the scale adopted in this figure, only a weak fluorescence K$\alpha$
line complex of oxygen at 0.52 --- 0.58~keV is visible. In Fig. 4, we 
plot the light curves of the energy flux in Fe~K$\alpha$ fluorescence 
line emission compared to the continuum fluxes in the same energy 
band and at hard X-rays (40 -- 50~keV). In agreement with our 
analytical estimate, the fluorescence line flux stays on a 
basically constant level until the time delay to the GRB 
equals the light travel time through the ionized medium. 
However, as is also obvious from Figs. 3 and 4, the expected 
flux of this long-duration line afterglow is very low compared 
to the prompt burst emission. For a GRB  located at $z = 1$, the 
luminosity shown in Fig. 4 translates into a flux of $F_{Fe \, K\alpha} 
\sim 6 \cdot 10^{-19}$~erg~cm$^{-2}$~s$^{-1}$ (where we adopted 
$H_0 = 75$~km~s$^{-1}$~Mpc$^{-1}$, $q_0 = 0.5$, $\Lambda = 0$). 

\begin{figure}
\epsfysize=6cm
\rotate[r]{
\epsffile[50 70 550 500]{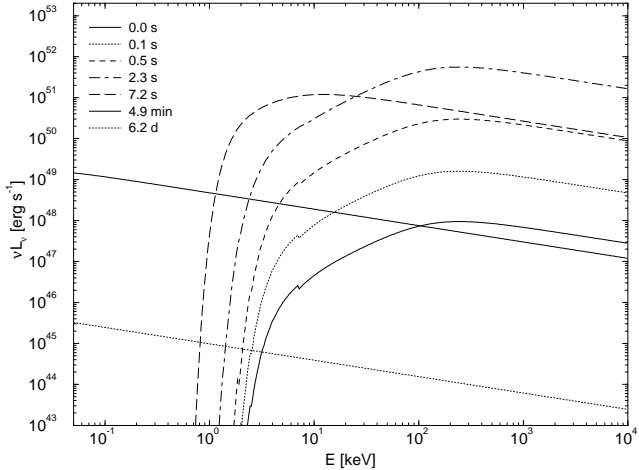}}
\caption[]{Temporal evolution of the observed GRB spectra at various 
times after the initial explosion. Parameters: $n_0 = 10^5$~cm$^{-3}$,
$r_{max} = 1$~pc, $\Gamma_0 = 150$, standard solar-system element 
abundances.}
\end{figure}

Figs. 5 and 6 illustrate the spectral evolution of a GRB in a
dense environment, more appropriate to a star-forming region.
Here, we assume a uniform external matter density of $n_0 =
10^5$~cm$^{-3}$, extended out to $r_{max} = 1$~pc. $\Gamma_0
= 150$ is used; all the other parameters are the same as for 
the previous example. Fig. 5 reveals that in this case the 
environment is photoionized rapidly because most of the 
absorbing material is very close to the GRB and therefore 
receives a very large ionizing flux, implying that it becomes 
optically thin to photoelectric absorption very quickly. This 
results in the rapid disappearance of the iron K absorption edge
after $\sim 1$~s. After $\sim 1.8$~min., the CBM becomes optically
thin to photoelectric absorption at virtually all frequencies.
The fluorescence line flux stays on a constant level over a 
timescale of one light travel time through the ionized region 
($\sim 3$~years in our example). At a redshift of $z = 1$, 
the iron K$\alpha$ fluorescence line flux is $F_{Fe \, K\alpha} 
\sim 3 \cdot 10^{-17}$~erg~cm~s$^{-1}$ over $\sim 5$~yr. Note 
that the times quoted in all figures refer to the reference 
frame at the redshift of the burst.

\begin{figure}
\epsfysize=6cm
\rotate[r]{
\epsffile[50 70 550 500]{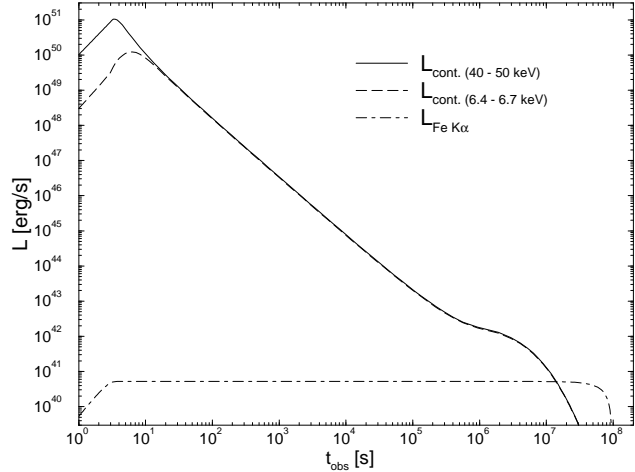}}
\caption[]{Energy flux light curves in the iron K$\alpha$ line band 
and in the hard X-ray continuum for the GRB spectral evolution shown 
in Fig. 4.}
\end{figure}

As in the case of optical absorption lines investigated by
Perna \& Loeb (1998), we find that due to
photoionization in a dense GRB environment any X-ray absorption
features (with the Fe~K edge being the most prominent one) 
will disappear rapidly after the onset of the GRB, on 
timescales which are anti-correlated with the external
matter density and positively correlated with the size
of the region over whith the CBM is extended. The detection 
of a varying Fe~K edge in GRB X-ray spectra therefore 
provides a means not only to measure the redshift of the 
GRB independently (only if the edge significantly varies, however,
can one be sure that the absorption happens in the CBM), 
but also to map the density structure of the GRB environment.
In Fig. 7 we have plotted the temporal evolution of the 
equivalent width 

\begin{equation}
EW = \int\limits_{E_K}^{\infty} dE \> \left( 1 - e^{-\tau_{abs} [E]}
\right)
\end{equation}
of the iron K edge for various combinations of the density 
and hydrogen column density (assuming standard solar-system 
abundances) of the CBM. The figure demonstrates
that the necessity to observe a time-varying Fe~K 
edge restricts this technique for measuring redshifts to fairly 
dense environments ($n \gtrsim 100$~cm$^{-3}$, $r_{max} \lesssim 
100$~pc) in which the evolution of the Fe~K edge is expected to 
be observable. Note that in any case the evolution of the Fe~K
edge happens on a timescale of $\lesssim 10$~s. It therefore 
appears to be subject to the same principal restrictions as the 
detection of optical absorption lines and requires rapid follow-up 
X-ray observations of $\gamma$-ray bursts and moderate spectral 
resolution at energies $\gtrsim 1$ keV during the prompt phase 
of the GRB.

\begin{figure}
\epsfysize=6cm
\rotate[r]{
\epsffile[50 70 550 500]{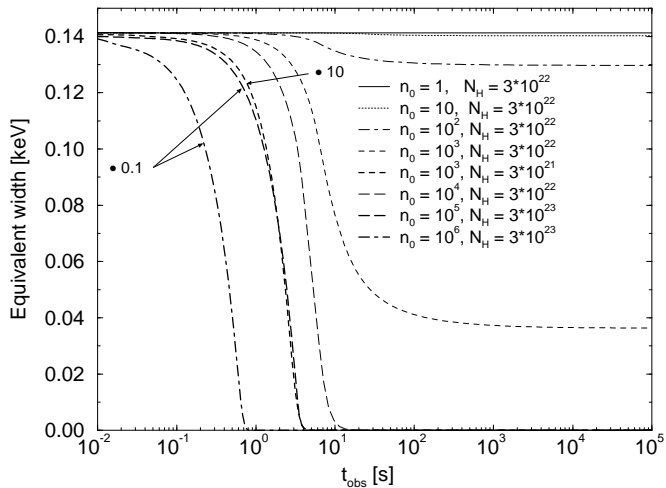}}
\caption[]{Temporal evolution of the equivalent width of the iron K edge for
different values of the CBM density ($n_0$ in units of cm$^{-3}$) and the 
hydrogen column density ($N_H$ in units of cm$^{-2}$). Standard GRB parameters 
and solar-system iron abundance ($X_{Fe} = 3.16 \cdot 10^{-5}$) are used.
The curve with $N_H = 3 \cdot 10^{21}$~cm$^{-2}$ is multiplied by a factor
of 10, the curves with $N_H = 3 \cdot 10^{23}$~cm$^{-2}$ are multiplied by
a factor 0.1 for clarity.}
\end{figure}

In contrast, the fluorescence Fe-line emission is observed 
on very long time scales. In order to follow 
fluorescence line afterglows of GRBs, no fast slewing of
an X-ray telescope is required or even helpful since they 
stay at a constant flux over timescales of months to years.
The fluorescence line flux only dominates over the continuum
flux from the burst after several months in the calculation shown.
However, we find that even for very dense GRB environments the
fluorescence line fluxes will hardly be detectable even  with upcoming
X-ray satellite missions such as AXAF or XMM: a flux of
$10^{-16}$~erg~cm$^{-2}$~s$^{-1}$ in the Fe~K$\alpha$ line 
(which is a very optimistic estimate), corresponding  to 
$\sim (1 + z) \cdot 10^{-8}$~photons~cm$^{-2}$~s$^{-1}$, 
observed with an instrument having an effective area of a few
$100$~cm$^2$ at the red-shifted line energy with an exposure 
of $100$~ksec would yield $\lesssim 1$~count. The fluorescence 
line fluxes are strongly limited by the fact that the hydrogen 
column density towards the GRB is restricted to $N_H \ll 
10^{24}$~cm$^{-2}$ since there is no evidence for Compton 
downscattering in the high-energy spectra and time profiles 
of GRBs.

The shape of the fluorescence line light curves we find in our 
simulations differs only slightly from the analytical results 
of Ghisellini et al. (1998), who find that the fluorescence 
line afterglow should decay as a power-law $F_{Fe K\alpha} 
(t) \propto t^{-0.1}$ after a maximum shortly after the burst.
We attribute this minor discrepancy to the fact that in the
analytical treatment of Ghisellini et al. (1998) 
the effect of photoionization on the GRB environment is neglected.
As mentioned earlier, the rapid photoionization of the CBM by
the leading light fronts of the burst radiation renders the 
environment optically thin to photoelectric absorption to the
later light fronts which therefore remain basically unabsorbed.
Thus, the illuminating fluence spectrum received by any point
in the CBM is basically the intrinsic, unabsorbed burst spectrum.

Perna \& Loeb (1998) used a temporal decay of the
afterglow emission of $F_{\nu} (t) \propto t^{-3/4}$, which
yields a rather strong late-time dependence of the variation
of the absorption features because most of the fluence from
the burst is emitted in the afterglow. Our choice of parameters
leads to a temporal decay with index $\gtrsim 1.1$, in agreement 
with observations, and a much less pronounced variation of the
absorption and fluorescence line features at late times (M\'esz\'aros
\& Rees 1998). In particular, most of the variation
in the Ly$\alpha$ absorption edge will occur on the time scale
on which the peak of the $\nu F_{\nu}$ spectrum is above the
Ly$\alpha$ edge energy, which is typically a few hours. This is
also the reason why the variation in the Fe~K edge equivalent
width as plotted in Fig. 7 occurs on very short time scales,
typically within the first few seconds of the burst.

The fluxes calculated here scale basically linearly with the
total burst energy $E_0$. The recent redshift determinations of
GRB~971214 (Kulkarni et al. 1998) and GRB~980703 
(Djorgovski et al. 1998b) seem to indicate that 
GRBs are at large cosmological distances, and their total energy 
release is in excess of $10^{53}$~erg. The choice of $E_0 = 
10^{54}$~erg, which we adopted in the two examples illustrated 
above, might therefore be representative of bright bursts.

\section{The low-energy slopes of GRB spectra}

In order to investigate whether the low-energy spectra of
GRBs are consistent with an optically thin synchrotron
spectrum depleted by photoelectric absorption, we have 
fitted time-resolved BATSE spectra of GRBs with a 
Band model (which is very similar to our analytical 
parametrization of the GRB spectrum), suffering photoelectric
absorption in a neutral medium where the opacity is
calculated using the model of Morrison and McCammon 
(1983). The asymptotic low-energy slope 
$\alpha$ is fixed to $-2/3$ (photon number index). For 
details of the fitting procedure, in the framework of 
an extensive study of the low-energy spectra of BATSE 
GRBs, see Crider et al. (1998).
One example of our fits to GRB~970111 using the Band model 
with photoelectric absorption is shown in Fig. 8. The fit 
has a reduced $\chi^2$ of 1.46, and there appear to be 
systematic deviations at photon energies below $\sim 100$~keV. 
Furthermore, there were 10 time bins for which no fit with
the photoelectric-absorption model was possible at all.

\begin{figure}
\epsfysize=7cm
\hskip 1cm \epsffile[50 200 550 680]{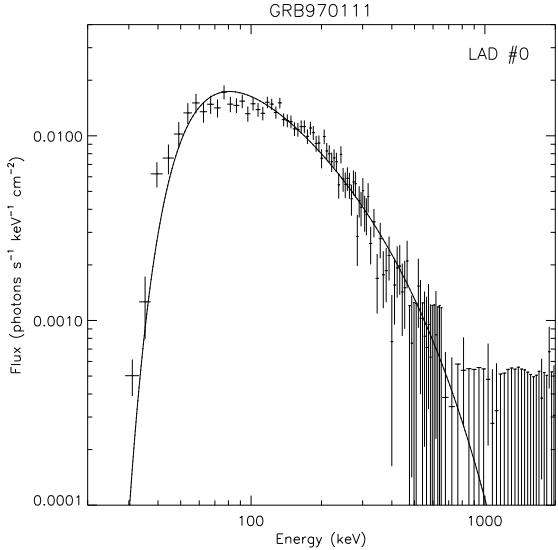}
\caption[]{Fit to a time-resolved BATSE spectrum of GRB~970111 using a
Band model with photoelectric absorption. The spectrum is integrated from
0.029 to 2.112~s. The fit results in a neutral hydrogen column of $N_H =
1.83 \cdot 10^{26} \, {\rm cm}^{-2}$ (assuming solar-system abundances),
corresponding to an absorption depth at the iron K edge of $\tau_K =
258$, and has a $\chi_{\nu}^2 = 1.46$.}
\end{figure}

Our fits using an underlying $F_{\nu} \propto \nu^{1/3}$
spectrum at low X-ray frequencies indicate that slightly 
lower values of $\tau_0$ are required in order to achieve 
acceptable fits to the data than in the situation 
investigated by Liang \& Kargatis (1994), 
where the intrinsic burst spectrum was assumed to be a 
straight power-law. However, the fits still require 
$\tau_0 \gtrsim 100$ for the first few seconds of
the burst. An example of the resulting $\tau_0$ values is
plotted in Fig. 9. Assuming a solar-system abundance of iron 
in the burst environment, this would correspond to a Thomson
depth of $\tau_T \gtrsim 100$. Therefore the CBM would be 
highly opaque to Thomson downscattering and would cause 
any short-term fluctuations in the intrinsic burst radiation 
to be smeared out over a typical timescale $\Delta t \sim 
\tau_T \, r_{max}/c$. Both of these consequences of such a 
high $\tau_T$ are in obvious contradiction to the 
observations. 

\begin{figure}
\epsfysize=6cm
\rotate[r]{
\epsffile[50 70 550 500]{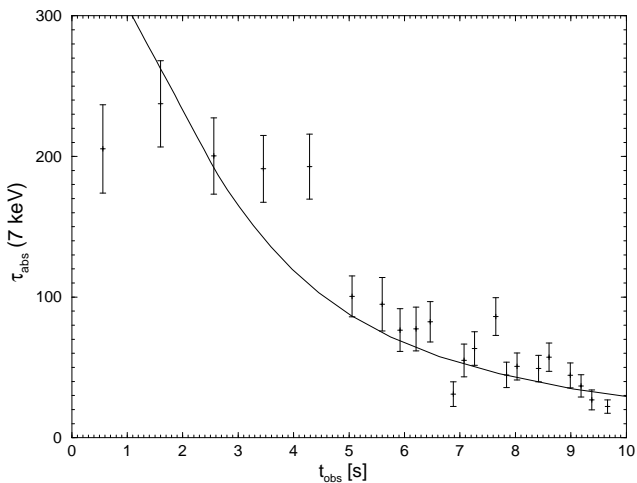}}
\caption[]{Photoelectric absorption opacity at the Fe~K-edge 
as a function of time. Parameters: $n_0 = 5 \cdot 10^6$~cm$^{-3}$, 
$r_{max} = 0.18$~pc, $\Gamma_0 = 100$, metal abundances = $100 \, 
\times$ solar-system metal abundances. Data points are the values 
obtained from fits to the time-resolved BATSE spectra of GRB~970111.}
\end{figure}

The above result suggests that the only way in which photoelectric
absorption could efficiently affect the low-energy slopes of GRB 
spectra up to $\sim 100$~keV would be a strong enhancement of the 
iron abundance in the CBM. This could happen as
the consequence of a metal enriched stellar wind produced
by the GRB progenitor (which could be a supermassive star, 
causing the GRB in a hypernova explosion, see Paczy\'nski
1998) and/or by massive stars in the vicinity
of the GRB, if it is located in a star-forming region. The
requirement that the Thomson depth of the CBM
be $\lesssim 1$ implies that the iron enrichment relative to
solar-system abundances in the burst environment would have to 
be of the order of 100. In order for the iron in the circumburst 
material to be completely photoionized within a few seconds, 
the metal-enriched matter must be highly concentrated around 
the burst location, $r_{max} \sim 0.2$~pc. For a metal
enrichment factor of 100, this implies an average density in
the immediate GRB environment of $n_0 \sim 5 \cdot 10^6$~cm$^{-3}$.
We did a model simulation adopting such parameters, which
roughly reproduces the temporal evolution of the photoionization
opacity as required by the fits to the time-resolved BATSE 
spectra of GRB~970111. This is illustrated in Fig. 9. 

\begin{figure}
\epsfysize=6cm
\rotate[r]{
\epsffile[50 70 550 500]{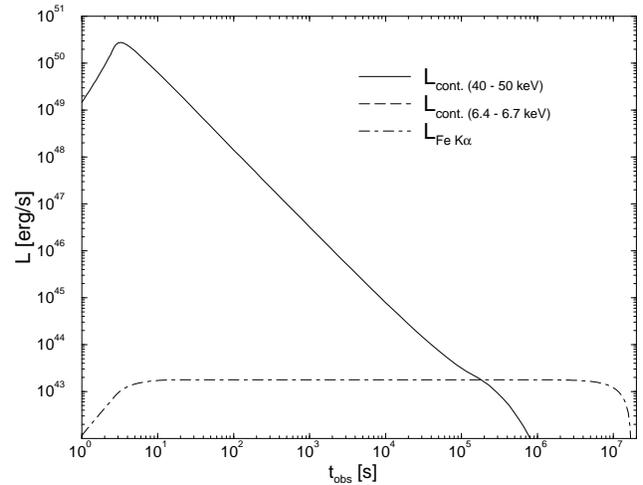}}
\caption[]{Energy flux light curves in iron K$\alpha$ lines 
and in two X-ray continuum bands resulting from the simulation 
reproducing the spectral evolution of GRB~970111 (Fig. 9). The
continuum flux at 6.4 - 6.7~keV is too heavily absorbed to be
visible on the scale of this plot.}
\end{figure}

Fig. 10 shows that this scenario would result in a 
considerable luminosity of delayed Fe~K$\alpha$ fluorescence
line emission, dominating over the X-ray afterglow continuum 
after $\sim 2 \, (1 + z)$~days, which at $z = 1$ 
would translate into a line flux of $F_{Fe K\alpha} \sim 
10^{-14}$~erg~cm$^{-2}$~s$^{-1}$. This flux is still 
in agreement with the non-detection of Fe~K$\alpha$ line 
emission in any GRB X-ray afterglow observed so far, even
those observed by ASCA (e. g., GRB~970228: Murakami et al.\ 
1997, GRB~970828: Yoshida et al. 1997). A Fe~K$\alpha$ line 
flux of $\sim 10^{-14}$~erg~cm$^{-2}$~s$^{-1}$, corresponding
to $\sim 2 \cdot 10^{-6}$~photons~cm$^{-2}$~s$^{-1}$
at a photon energy of $\sim 3.2$~keV, would even be hard 
to detect, e. g., with the ACIS-S detector on board the 
AXAF satellite within a reasonable exposure time 
($\lesssim 100$~ksec).

Thus, from the analysis of photoelectric absorption, photoionization
and fluorescence line emission alone, the idea of the hard low-energy
slopes of time-resolved BATSE GRB spectra being the result of
photoelectric absorption in the CBM cannot be formally ruled out. 
However, the required extremely high metal enrichment and the
peculiar spatial distribution necessary to reproduce the best-fit
absorption depths look like fine-tuning and seem unlikely to be 
produced in a natural way. Furthermore, the fits to the time-resolved
BATSE spectra of GRB~970111 with the photoelectric-absorption model
were at best marginal. The BeppoSAX data on GRB~970111 (Feroci 
et al. 1998) constrain the photoelectric-absorption 
model severely, depending on the (unknown) redshift of this burst. 
In the fits presented above, the absorber was assumed to be located
at $z = 0$. If the redshift of GRB~970111 is $z \gtrsim 1$, implying 
that the iron K edge is near or below the low-energy end of the
BeppoSAX WFC energy range, then the 2 --- 10~keV flux would have 
been heavily absorbed and thus undetectable by the WFC. Crider et 
al. (1998) show that in this case the combination 
of time-resolved BATSE spectra with simultaneous BeppoSAX data 
indicates that the photoelectric-absorption model is most likely 
inconsistent with the broadband X-ray spectrum of GRB~970111.

\section{Summary and conclusions}

We have analyzed the effects of photoionization, photoelectric
absorption and fluorescence line emission in the vicinity of cosmological
$\gamma$-ray bursts. Under the assumptions that the GRB emission is
isotropic and GRBs are hosted in galaxies, we have calculated the 
expected time-dependence of photoelectric absorption features in 
the X-ray afterglows of $\gamma$-ray bursts and the flux and light
curves of fluorescence lines from the photoionized CBM.

We find that an Fe~K edge is the most easily detectable
signature of the CBM in the X-ray continuum afterglows of
GRBs. If the effect of photoionization on the CBM is 
identified by virtue of a reduction of the optical depth 
at the iron K edge as the burst evolves, this absorption 
can be attributed to material in the immediate vicinity 
of the GRB and may serve as an independent redshift indicator, 
as well as a method to map the GRB environment. This method 
is applicable to a dense environments such as
star-forming regions. In more dilute, extended environments, 
the flux of the GRB emission becomes too weak at large
distances from the burst location to be efficient in terms
of photoionization.

We find that the ionization sphere, out to which photoionization
and subsequent fluorescence line emission are efficient, is 
typically determined by the size of the region of considerable 
density around the GRB. This implies that the fluorescence line 
flux will remain relatively constant over
timescales of months to years. This implies that 
rapid follow-up observations of $\gamma$-ray bursts are not needed to
search for fluorescence lines. Instead, if there is any chance at all to
detect fluorescence line afterglows, the candidate host galaxies of GRBs
should be monitored with a sensitive X-ray pointing instrument such
as AXAF over weeks to years after the GRB.

The depth of the Fe~K absorption edge yields a measure of
the Fe column density of the CBM, while the rate of decrease
of the absorption edge (if observable) yields an estimate of
the average distance of the CBM from the burster, thus allowing
an estimate of the density structure of the GRB environment.
The level of fluorescence line flux relative to the peak flux
of the GRB continuum is also related to the average density of 
the CBM. While most of the parameters determining the evolution 
of a GRB blastwave in the framework of the cosmological blastwave 
model may be deduced from the flux and fluence spectra and spectral 
evolution of the prompt GRB emission and the continuum afterglow 
(if the redshift of the source is known), the density of the 
CBM always remained a free parameter. 
The absorption edges and the fluorescence line afterglow 
predicted in this paper provide a method to map the 
density of material in the GRB environment. Observation
of vanishing absorption edges and/or fluorescence lines 
following a GRB also provides a very accurate measure of 
the redshift of the GRB. If the redshift of the host galaxy 
could be measured independently, this could confirm or rule 
out the yet unproven claim that those GRBs which are spatially
coincident with host galaxies (e. g., GRB~970508: Bloom et al.
1998, GRB~971214: Kulkarni et al. 1998, GRB~980613: Djorgovski 
et al. 1998a, GRB~980703: Djorgovski et al. 1998b) are actually 
located within the proposed host galaxy. 

Our prediction of the flux level of the fluorescence line 
afterglows is based on the assumptions that
(1) the respective GRB is located within the proposed host 
galaxy, and (2) the GRB emission is isotropic. 
If the latter assumption is not true, only a small solid angle 
of the CBM is photoionized by the GRB radiation,
which is why only  a comparatively small volume can contribute 
to the fluorescence line afterglow. Furthermore, since this volume 
is exclusively located at small angles to our line of sight, only 
small time delays result. Therefore, the fluorescence line emission 
will be overwhelmed by the bright continuum afterglow.

We have re-analyzed the possibility of photoelectric absorption
causing the hard low-energy slopes of time-resolved BATSE GRB
spectra. The fits using this spectral form are at best marginal,
and the density structure around the GRB needs to be fine-tuned
in a rather unnatural way in order to reproduce the observed 
spectral evolution of the low-energy spectrum of GRB~970111. We
therefore confirm earlier findings that alternative mechanisms
are more likely to be responsible for the hard low-energy snapshot
spectra of GRBs and their spectral evolution.

\acknowledgements{We wish to thank Jon C. Weisheit and Tim Kallman
for helpful discussions. This work was partially supported by
NASA grant NAG~5-4055.}

\end{document}